\documentclass[11pt,tightenlines,eqsecnum,floats,aps,prd,showpacs,amsmath,superscriptaddress,amssymb,nofootinbib,longbibliography]{revtex4-1}

\usepackage{bm}
\usepackage[latin1]{inputenc}
\usepackage[spanish,english]{babel}
\usepackage{amsfonts}
\usepackage{amssymb}
\usepackage{graphicx}
\usepackage{color}
\usepackage{stmaryrd}
\usepackage{mathtools,slashed}

\usepackage{amsmath}
\usepackage{soul}
\usepackage[normalem]{ulem}

\newcommand{\be}{\begin{equation}}
\newcommand{\ee}{\end{equation}}
\newcommand{\bea}{\begin{eqnarray}}
\newcommand{\eea}{\end{eqnarray}}
\newcommand{\ben}{\begin{enumerate}}
\newcommand{\een}{\end{enumerate}}

\begin{document}

\title{{\bf A  Non-Local Action for Electrodynamics: \\
Duality Symmetry and the Aharonov-Bohm Effect,~Revisited\footnote{Article prepared for  the Special Issue: ``Symmetry in Electromagnetism'', published in {\it Symmetry}\  11 (2019) no.10, 1191.
}}}

\author{Joan Bernabeu}\email{Joan.Bernabeu@physik.uni-muenchen.de}\affiliation{Physik Department, Ludwig-Maximilians-Universität München, Theresienstraße 37, D-80333 München, Germany;}
\affiliation{Physik Department, Technische Universität München, James Franck Straße 1, D-85748 Garching, Germany;}
\author{Jose Navarro-Salas}\email{jnavarro@ific.uv.es}
\affiliation{Departamento de Fisica Teorica, IFIC. Centro Mixto Universidad de Valencia - CSIC.  Facultad de Fisica, Universitad de Valencia, Burjassot-46100, Spain.}

\date{\today}

\begin{abstract}

{A non-local action functional for electrodynamics depending  on the electric and magnetic fields, instead of potentials, has been proposed in the literature. In this work we elaborate and improve this proposal. We also use 
this formalism to confront the electric-magnetic duality symmetry of the electromagnetic field and the Aharonov--Bohm effect, two subtle aspects of electrodynamics that we examine in a novel way. We show how the former can be derived from the simple harmonic oscillator character of vacuum electrodynamics, while also demonstrating how the magnetic version of the latter
naturally arises in an explicitly non-local manner.}


\end{abstract}


\maketitle

\section{Introduction}

Locality is a preferred virtue of fundamental field theories. Electrodynamics, the paradigm of field theory, and general relativity, the modern and finest description of gravity, are very important examples. Both theories are consistent with local causality and the conservation of energy and momentum. Maxwell's and Einstein's equations are systems of partial differential equations for their fundamental fields: the  electromagnetic and metric tensors, respectively. The two sets of field equations 
can also be derived from an action functional. The Hilbert-Einstein action itself is also local in the metric field. However, to derive the  Maxwell equations from a local action one has to introduce the electromagnetic potentials. To construct an action depending exclusively on gauge invariant quantities  
one must necessarily sacrifice   locality. This issue is very rarely treated in the literature, despite of the fact that it is a question that may naturally arise in graduate courses on basic field theory and classical electrodynamics (see, for instance \cite{DeWitt62, ColemanAspects} and references therein).
 Within the context of constrained dynamical systems \cite{faddeev-jackiw, Henneaux-Teitelboim92, Barbero19}, a non-local action functional describing  Maxwell theory, dependent  on the electric and magnetic fields, was sketched in Ref. \cite{jackiw1993constrained}.  In this paper we will focus on this proposal and related  aspects of quantum mechanics and the theory of Noether's symmetries. 

As remarked above, electrodynamics is commonly formulated in terms of Hamilton's variational principle through the action functional
$S[A^\mu] = \int d^4 x \mathcal{L}_\text{EM}$, where the Lagrangian density for the electromagnetic field in the presence of an external current source $J^\mu\equiv (\rho, \mathbf{J})$, is given by \cite{jacksonbook, qftbook}
\be\label{standardLagrangian}
\mathcal{L}_\text{EM} \equiv - \frac{1}{4}F_{\mu\nu}F^{\mu\nu} - A^\mu J_\mu \ . 
\ee
The action is regarded as a  functional  of the 4-vector potential  $A^\mu = (A^0,\mathbf{A})$, where  $F^{\mu\nu} \equiv \partial^\mu A^\nu - \partial^\nu A^\mu$ is the electromagnetic field tensor. $E^i= -F^{0i}$ and $B^i= -\frac{1}{2} \epsilon^{ijk} F^{jk}$ are the components of the electric and magnetic fields ($\mathbf{E}$ and $\mathbf{B}$), respectively, and the metric $\eta = \text{diag}(1,-1,-1,-1)$ was used to lower and raise indices in $J_\mu$, $F_{\mu\nu}$, and $\partial^\mu$ (e.g., $J_\mu = \eta_{\mu\nu} J^\nu$). 
[ Throughout this work we use Lorentz-Heaviside units and take $c=1$.  We also assume the Einstein summation convention for repeated indices and $\epsilon^{123}=1$. Additionally, greek letter indices refer to time and Cartesian space coordinates  whereas latin letter indices only refer to the latter. Furthermore, simultaneous spacetime points are labelled as $x \equiv (t,\mathbf{x})$ and $x' \equiv (t, \mathbf{x}')$. Finally, it is assumed that all fields decay to 0 at~infinity.]

The inhomogeneous Maxwell equations 
\begin{equation}\label{ampere}\nabla \times \mathbf{B}  - \partial_t \mathbf{E} = \mathbf{J} \ , \end{equation} 
\begin{equation}\label{gauss}
\nabla\cdot\mathbf{E} =\rho \ , 
\end{equation} 
are obtained by varying the action with respect to $\delta A^\mu$ and imposing $\delta S =0$.
One gets immediately $\partial_\mu (\partial^\mu A^\nu - \partial^\nu A^\mu)= J^\nu$, and rewriting the potential in terms of the electric and magnetic fields, Gauss'~law (\ref{gauss}) and the Ampere-Maxwell equation (\ref{ampere}) are readily obtained. 
The fact that (\ref{ampere}) and (\ref{gauss}) only hold on-shell (i.e., when the Euler-Lagrange equations for $A^\mu$ hold) contrasts  with the off-shell nature of the homogeneous Maxwell equations
\begin{equation}\label{Mh1}\nabla \times \mathbf{E} + \partial_t \mathbf{B} =0 \ , \end{equation}
\begin{equation} \label{Mh2} \nabla \cdot \mathbf{B} =0 \ , \end{equation}
which are trivially satisfied by the definition of $F^{\mu\nu}$ in terms of the potentials, 
 or equivalently $\mathbf{E} = - \nabla A^0 - \frac{\partial}{\partial t} \mathbf{A}, \ \mathbf{B} = \nabla \times \mathbf{A}$ in vector notation.
 This distinction between two types of Maxwell equations can seem somewhat forced, as in essence it is only due to the choice of $A^\mu$ as the 
 field of the action functional.
  Nevertheless, it is the price to be paid to deal with a {\it local} action, i.e., one where $\mathcal{L}_\text{EM}$ depends on the value of  $ A^\mu(x)$ and finitely many derivatives at a
single spacetime point $x$.

An alternative local action functional is given by \cite{ColemanAspects}
\be\label{standardLagrangian2}
 S[A_\mu, F_{\mu\nu}] = \int d^4x[ \frac{1}{4}F_{\mu\nu}F^{\mu\nu} -\frac{1}{2} F^{\mu\nu}(\partial_\mu A_\nu - \partial_\nu A_\mu) - A^\mu J_\mu] \ . 
\ee
$F_{\mu\nu}$ and $A_\mu$ are here considered to be completely independent dynamical variables. The equation of motion for $F_{\mu\nu}$ is $F_{\mu\nu}= \partial_\mu A_\nu - \partial_\nu A_\mu$, and plugging this into the action (\ref{standardLagrangian2}) one gets the standard action $S[A^\mu] = \int d^4 x \mathcal{L}_\text{EM}$. This alternative first-order action (\ref{standardLagrangian2}) is very efficient to prove \cite{ColemanAspects} that the  covariant Feynman rules for quantum electrodynamics obtained from the functional integral approach are indeed equivalent to the rules derived within the canonical formalism.


 The use of potentials in (\ref{standardLagrangian}) is also useful to study electrodynamics with matter sources. 
Recycling~the field-matter interaction term $-A^\mu J_\mu$ present in (\ref{standardLagrangian}), inserting the charge distribution (the dot refers to a total time derivative)
\be\label{particleDistribution}
\rho(x') = e\delta^3(\mathbf{x}(t)-\mathbf{x}') \quad \text{and} \quad \mathbf{J} (x')= e\dot{\mathbf{x}}(t)\delta^3(\mathbf{x}(t)-\mathbf{x}'),
\ee
and adding a kinetic energy term, the standard Lagrangian that describes the motion of a non-relativistic particle of mass $m$ and charge $e$ within an external electromagnetic field,
\be\label{standardParticleLagrangian}
L_\text{p} = \frac{1}{2}m\dot{\mathbf{x}}^2 + e\mathbf{A}\cdot\dot{\mathbf{x}} - eA^0,
\ee
is recovered. Despite the fact that the action $S_\text{p}[\mathbf{x}] = \int dt L_\text{p} $ is explicitly dependent on the potentials, the equations of motion, which in this case are just the Lorentz force
\be\label{lorentzForce}
m\ddot{\mathbf{x}} = e(\mathbf{E} + \dot{\mathbf{x}}\times\mathbf{B}),
\ee
can be expressed solely
in terms of the electromagnetic field, similarly to the case of Equations (\ref{ampere}) and~(\ref{gauss}) with respect to the action $S$. Consequently, in classical mechanics where $\delta S_\text{p} = 0$ strictly defines the dynamics of the particle, this formulation does not pose anything more than possibly an aesthetic nuisance. However, in the context of quantum mechanics, where the contribution of trajectories with $\delta S_p \neq 0$ to the path integral is not negligible \cite{feynman1965quantum}, this formulation does become an issue with the interpretation of the Aharonov--Bohm (AB) effect   \cite{aharonov1959significance, aharonov1961further, feynman1979feynman, abbook, sakurai2017modern} . 

As mentioned above, the first aim of this paper is to  study the non-local formulation  suggested by Jackiw \cite{jackiw1993constrained}. It is of first-order in time derivatives, but spatially non-local. We will elaborate on this proposal finding a slightly more simplified expression for the action functional  
than that originally proposed \cite{jackiw1993constrained} (see the comments after Equation (\ref{constraints})).  This alternative non-local action  turns out to be very efficient to analyze the electric-magnetic duality symmetry of free electrodynamics, and, as a bonus,  to gain new insights on the AB effect.\\


\section{The Free Non-Local (Duality Invariant) Action\label{nonlocalfree}
}

A wide family of first-order Lagrangians in classical mechanics can be expressed as
\be\label{classic1Lagrangian}
L= \omega_{ij}\dot{q}^ip^j - H(q,p) \ , 
\ee
where the constants $\omega_{ij}$ are the components of the off-diagonal block term of the symplectic tensor
\be
\Omega = \begin{pmatrix}
0 &\omega \\
-\omega & 0
\end{pmatrix}
\ee
and $H(q,p)$ is the system's Hamiltonian \cite{jackiw1993constrained, faddeev-jackiw}. As the notation hints, $q = \{q^i\}$ and $p = \{ p^i\}$ are the sets of (phase space) variables. If $\omega$ has an inverse $\omega^{-1}$, then their brackets are simply $ \{q^i,p^j\} \equiv \omega^{ij}$ ($\{q^i, q^j\} = \{p^i, p^j\} = 0$), where $\omega^{ij}$ are the components of $\omega^{-1}$. The conventional choice for simple Hamiltonian systems is $\omega_{ij}= \delta_{ij}$, and hence $q$ and $p$ are canonically conjugate variables with $\{q^i,p^j\} = \delta^{ij}$. However, when $\omega$ is not invertible, one typically faces a constrained system, examples of which we give below. 

The Lagrangian (\ref{classic1Lagrangian}) can be generalized to a Lagrangian density for the context of field theory. Besides summing over the discrete degrees of freedom in the non-Hamiltonian component of (\ref{classic1Lagrangian}), one~must also sum over (i.e., integrate) the continuous degrees of freedom. Thus, the Lagrangian density of the conjugate fields $\phi$ and $\pi$ can be expressed in terms of the Hamiltonian density $\mathcal{H}(\phi,\pi)$~as
\be\label{field1Lagrangian}
\mathcal{L} = \int d^3x' \omega_{ij}(\mathbf{x},\mathbf{x}')\partial_t\phi^i(x)\pi^j(x') - \mathcal{H}(\phi,\pi)
\ee
with  $\{\phi^i(x),\pi^j(x')\} \equiv \omega^{ij}(\mathbf{x},\mathbf{x}')$, if $\omega$ is invertible. The most conventional choice for $\omega$ in field theory is
$\omega_{ij}(\mathbf{x},\mathbf{x}')= \delta_{ij} \delta^3(\mathbf{x}-\mathbf{x}')$,
which leads to the local Lagrangian density
$\mathcal{L} =\partial_t\phi^i(x)\pi^j(x) - \mathcal{H}(\phi,\pi)$.~
For 
\be\mathcal{H}(\phi,\pi)= \frac{1}{2} (\pi^2 + (\nabla \phi)^2) + \frac{1}{2}m^2\phi^2\ee
we have the usual free scalar Klein-Gordon theory, with field equations
$\partial_t \phi = \pi$ and $\partial_t \pi = (\nabla^2 - m^2)\phi$,
which easily combine into the Klein-Gordon wave equation $(\partial_t^2 -\nabla^2 + m^2) \phi=0$, consistent with $\{\phi^i(x),\pi^j(x')\} = \delta^{ij} \delta^3(\mathbf{x}-\mathbf{x}')$. 


A more involved  example is given by taking $\omega_{ij}(\mathbf{x},\mathbf{x}')$ as the divergenceless or transverse delta~function 
\be\label{transverseDelta}
\omega_{ij}(\mathbf{x},\mathbf{x}')= \delta^\text{T}_{ij}(\mathbf{x}-\mathbf{x}') \equiv \delta_{ij}\delta^3(\mathbf{x}-\mathbf{x}') + \partial_i \partial_j \frac{1}{4\pi|\mathbf{x}- \mathbf{x}'|} \ .
\ee

It is convenient to briefly recall here that a generic vector field $\mathbf{F}$ always decomposes univocally~\cite{griffiths1981introduction} into a transverse vector $\mathbf{F}_\text{T}$, obeying $\nabla \cdot \mathbf{F}_\text{T}=0$, plus a longitudinal one  $\mathbf{F}_\text{L}$, with $\nabla \times \mathbf{F}_\text{L}=0$. The~transverse delta can then be used to project the transverse component,
\be \int d^3x'  \delta^\text{T}_{ij}(\mathbf{x}-\mathbf{x}')F^j(\mathbf{x}') = F^i_\text{T}(\mathbf{x}) \ . \ee

Choosing the variables to be vector fields $\phi \to \mathbf{E}$, $\pi \to \mathbf{A}$ with a Hamiltonian density given by
\be \mathcal{H}_0(\mathbf{E},\mathbf{A}) = \frac{1}{2}[(\mathbf{E}^2 + (\nabla\times\mathbf{A})^2] \ , \ee
then the (non-local) Lagrangian density reads
\be\label{EandA}
\mathcal{L}_0 = \int d^3x'  \delta^\text{T}_{ij}(\mathbf{x}-\mathbf{x}')\partial_tE^i(x) A^j(x') - \mathcal{H}_0(\mathbf{E},\mathbf{A}) \ .
\ee
In contrast with the Klein-Gordon example, this Lagrangian density, due to the extra contribution to the delta function,  cannot be reduced to a local one in terms of the chosen fields $\mathbf{E}, \mathbf{A}$. Furthermore,~(\ref{EandA}) is invariant under gauge transformations $\mathbf{A}' = \mathbf{A} + \nabla \xi$. By taking variations and assuming the appropriate boundary conditions one obtains the field equations
\be E^i = - \int d^3x'  \delta^\text{T}_{ij}(\mathbf{x}-\mathbf{x}')\partial_t A^j(x') = - \partial_t A_\text{T}^i \ , \ee
\be [\nabla \times (\nabla \times \mathbf{A})]^i = \int d^3x'  \delta^\text{T}_{ij}(\mathbf{x}-\mathbf{x}')\partial_t E^j(x')  = \partial_t E_\text{T}^i \ . \ee

However, after some manipulations one can transform the above equations into the following set of local field equations
\be \partial_t\mathbf{E}= \nabla \times (\nabla \times \mathbf{A}) \ , \ \ \ \ \ \nabla \times \mathbf{E} + \partial_t( \nabla \times \mathbf{A}) =0 \ , \ee
\be\label{gaussfree} \nabla \cdot \mathbf{E}=0 \ . \ \ \ \ \ \  \ee
The source-free versions of (\ref{ampere})--(\ref{Mh1}) are recovered with the identification $\mathbf{B} = \nabla \times \mathbf{A}$. Equation (\ref{Mh2}) identically follows from the definition of the magnetic field in terms of $\mathbf{A}$, hence completing the full set of vacuum Maxwell equations. Note how the Gauss law constraint (\ref{gaussfree}) was obtained without explicitly introducing any Lagrange multiplier. Also note how the transverse delta can project $\mathbf{A}_\text{T}$, leading to the  Lagrangian density
\be\label{EandAT}
\mathcal{L}_0 = \partial_t\mathbf{E}\cdot\mathbf{A}_\text{T} - \frac{1}{2}\left[\mathbf{E}^2 + (\mathbf{\nabla}\times\mathbf{A}_\text{T})^2\right] \ , 
\ee
where the longitudinal component of $\mathbf{A}$ has naturally decoupled from the theory.  That this is the case seems natural, as $\mathbf{A}_\text{L}$ does not possess indispensable physical value due to the aforementioned gauge invariance.  Please note that although  (\ref{EandAT}) is apparently a local expression, there is a hidden non-locality in the (constrained and gauge-independent) transverse vector potential.  Solving now the constraint (\ref{gaussfree}) (i.e., taking $\mathbf{E} = \mathbf{E}_\text{T}$) into (\ref{EandA2}) we finally get
 \be\label{EandA2}
\mathcal{L}_0 = \partial_t\mathbf{E}_\text{T}\cdot\mathbf{A}_\text{T} - \frac{1}{2}[\mathbf{E}_\text{T}^2 + (\nabla\times\mathbf{A}_\text{T})^2] \ .
\ee

In this way we therefore recover the completely reduced form of the electromagnetic Lagrangian density. 
A bonus of the above discussion is that one can immediately work out the brackets of the theory: $\delta^\text{T}_{
ij}(\mathbf{x}-\mathbf{x}')$ can be inverted for transverse vector fields and hence the expected \cite{Bjorken,vogel2006quantum,weinberg2015lectures} $ \{E_\text{T}^i(x),A_\text{T}^j(x')\} = \delta^{\text{T} ij}(\mathbf{x}-\mathbf{x}') $ is derived.

\subsection{Non-Local Formulation for the Electromagnetic Field in Terms of $\mathbf{E}$ and $\mathbf{B}$}
Our last and most important example consists of defining the object $\omega_{ij}(\mathbf{x},\mathbf{x}')$ for the electric and magnetic field themselves. The solution involves a  derivative of the Green's function for the Laplacian operator $ \nabla^2 \equiv \partial_i\partial_i$, and it is given by
\begin{equation}\label{symplectic} 
\omega_{ij}(\mathbf{x},\mathbf{x}') = \epsilon^{ijk}\partial_k \frac{-1}{4\pi|\mathbf{x}- \mathbf{x}'|} \ .
\end{equation} 

This expression can be regarded as the simplest way to enforce the appropriate physical dimensions for $\omega_{ij}(\mathbf{x},\mathbf{x}')\partial_t E^i B^j$  and consistency with respect to electric-magnetic duality symmetry (see next subsection for more details). Together with the conventional electromagnetic Hamiltonian density we can construct, in the absence of sources, the action $S_\text{NL,0}[\mathbf{E},\mathbf{B}] = \int d^4x \mathcal{L}_\text{NL,0}$, a functional exclusively dependent on the electromagnetic field, with a first-order Lagrangian density
\begin{equation}\label{nonlocalFreeLagrangian}
\mathcal{L}_{\text{NL,0}}=\int d^3x' \ \omega_{ij}(\mathbf{x},\mathbf{x} ')\partial_tE^i(x)B^j(x') - \frac{1}{2}(\mathbf{E}^2(x) + \mathbf{B}^2(x))\ .
\end{equation}

It is quite remarkable that this action 
 yields all of the four vacuum Maxwell equations. The~integral term  in (\ref{nonlocalFreeLagrangian})   introduces an explicit non-locality, as the fields at spatially separated points $x = (t,\mathbf{x})$ and $x' = (t,\mathbf{x}')$ "interact'' with one another. This coupling is nonetheless weighed by $\omega_{ij}(\mathbf{x},\mathbf{x}')$, leading~it to steadily decay as $\mathbf{x}$ and $\mathbf{x}'$ become further apart.
Taking variations of $E^i$ and $B^i$, simultaneously~exploiting the standard fall-off conditions of the fields at infinity, one can show that the equations of motion are just the Hemholtz decomposition \cite{griffiths1981introduction} of the free electromagnetic field,
\begin{equation}\label{deltaEvac}
E^i (x)= -\int d^3x' \omega_{ij}(\mathbf{x},\mathbf{x}') \partial_t B^j (x') \ , 
\end{equation}
\begin{equation}\label{deltaBvac}
B^i(x) = \int d^3x' \omega_{ij}(\mathbf{x},\mathbf{x}') \partial_t E^j (x') \ .
\end{equation}
Applying a divergence and a curl on (\ref{deltaEvac}) and (\ref{deltaBvac}) immediately provides the vacuum versions of Equations (\ref{ampere})--(\ref{Mh2}), 
\be \nabla \times \mathbf{E} = -\partial_t \mathbf{B} \ , \quad \nabla \times \mathbf{B} = \partial_t \mathbf{E} \ , \ee
\be\label{constraints} \nabla \cdot \mathbf{E} = 0 \ , \quad \nabla \cdot \mathbf{B} = 0 \ . \ee

The non-local Lagrangian density  $\mathcal{L}_{\text{NL,0}}$  is similar to the one given in Ref. \cite{jackiw1993constrained}, up to the contributions of two Lagrange multipliers, which we find unnecessary in the absence of sources.  As in the previous case [(\ref{EandA}) and (\ref{EandA2})], the constraints (\ref{constraints}) can be solved into the Lagrangian density~(\ref{nonlocalFreeLagrangian}). In~this situation, where the fields are necessarily transverse, $\omega$ does possess an inverse, leading to the anticipated \cite{vogel2006quantum} brackets
\be\label{EBbracketANDHam}
\{E_{\text{T}}^i(x),B_{\text{T}}^j(x')\} = -\epsilon^{ijk}\partial_k\delta^3(\mathbf{x}-\mathbf{x}') \ . \ee

Note also how (\ref{EandA}), and consequently (\ref{EandA2}), can also be recovered from (\ref{nonlocalFreeLagrangian}) by introducing the vector potential $\mathbf{A}$ such that $\mathbf{B} = \nabla \times \mathbf{A}$. 

\subsection{{Electric-Magnetic Duality  Symmetry}}

The fact that (\ref{nonlocalFreeLagrangian}) is formulated solely in terms of $\mathbf{E}$ and $\mathbf{B}$ means that it is manifestly dual, quite in contrast to the standard formulation (\ref{standardLagrangian}).  It is straightforward to prove that the discrete transformations $\mathbf{E} \rightarrow -\mathbf{B}, \ \mathbf{B} \rightarrow \mathbf{E}$ and their continuous generalization as electric-magnetic duality rotations \cite{jacksonbook} with parameter $\theta$,
\begin{equation}\label{dualMatrix}
\begin{pmatrix}
\mathbf{E}' \\
\mathbf{B}'
\end{pmatrix}
=
\begin{pmatrix}
\cos\theta & \sin\theta \\
-\sin\theta & \cos\theta
\end{pmatrix}
\begin{pmatrix}
\mathbf{E} \\
\mathbf{B}
\end{pmatrix},
\end{equation}
leave the Maxwell equations invariant. It is, however, not such a simple task \cite{calkin1965invariance, deser1976duality, deser1982duality, agullo2017electromagnetic} to prove that (\ref{dualMatrix}) are a symmetry in the Noether sense, i.e., that their infinitesimal version
\begin{equation}\label{infidual}
\delta \mathbf{E} = \theta \mathbf{B}, \qquad \delta \mathbf{B} = -\theta \mathbf{E},
\end{equation}
leaves the Lagrangian $L = \int d^3 \mathcal{L}$ invariant, up to a total time derivative and without making use of the field equations.  

Employing the standard formulation (\ref{standardLagrangian}), the transformations (\ref{infidual}) clearly will not suffice as Noether's theorem requires the transformations of the dynamic fields, $A^\mu$ in this case. However,~the~problem is actually deeper. The introduction of the potentials implies that Equations~(\ref{Mh1})~and~(\ref{Mh2}) hold, which for consistency would also require, through the use of (\ref{infidual}), the~equations $\nabla \times \mathbf{B}  - \partial_t \mathbf{E} =0$ and $\nabla\cdot \mathbf{E}=0$. However, within the Lagrangian formalism it is forbidden to use the latter (on-shell) equations to prove that the duality rotations are a symmetry of the theory.  Consequently, the transformation in (\ref{infidual}) cannot be applied directly \cite{deser1976duality, deser1982duality} on (\ref{standardLagrangian}) with Noether's Theorem. A way out of this tension is to project the  original duality rotations on the transverse fields $( \mathbf{E}_\text{T}, \ \mathbf{A}_\text{T})$ and consider the reduced  Lagrangian (\ref{EandA2}) \cite{deser1976duality, deser1982duality}. The new form of the duality  symmetry is then non-local. 

On the other hand, the application of Noether's theorem with (\ref{nonlocalFreeLagrangian}) is swift and even elegant. While~the bracket has become more intricate in the transition from using $\mathbf{A}$ and $\mathbf{E}$ to $\mathbf{B}$ and $\mathbf{E}$, the~Hamiltonian density now has the well known form of the isotropic simple harmonic oscillator~(SHO),
\be
H(q,p) = \frac{1}{2}(q^2 + p^2) \quad \text{(normalized)}.
\ee 

The presence of the SHO in this context shouldn't be too surprising, as it is a well-known fact that vacuum electromagnetic field satisfies the wave equations $\partial_\mu\partial^\mu \mathbf{E}$ and $\partial_\mu\partial^\mu \mathbf{B} = 0 $, which are just the field version of the SHO equations $\ddot{q}^i + k^2q^i = 0$ and $\ddot{p}^i + k^2p^i = 0$. Thus, (\ref{nonlocalFreeLagrangian}) can be viewed as a the first-order Lagrangian of a SHO with \textit{non-canonical}, i.e., $\{q^i,p^j\} \neq \delta^{ij}$, commutation relations.
As with the \textit{canonical}, i.e., $\{q^i,p^j\} = \delta^{ij}$, SHO, this system is also invariant under phase space~rotations
\begin{equation}\label{phaseSpaceRotation}
\begin{pmatrix}
q'^i   \\
p'^i
\end{pmatrix}
=
\begin{pmatrix}
\cos\theta & \sin\theta \\
-\sin\theta & \cos\theta
\end{pmatrix}
\begin{pmatrix}
q^i\\
p^i
\end{pmatrix}.
\end{equation}

However, while in the canonical case this symmetry implies conservation of energy, the non-trivial case preserves a more general quantity, which using Noether's theorem is straight-forwardly shown to~be
\be
Q = \frac{1}{2}\omega_{ij}(q^iq^j + p^ip^j).
\ee

Of course, phase space rotations (\ref{phaseSpaceRotation}) are just electric-magnetic rotations (\ref{dualMatrix}) in the formalism of~(\ref{nonlocalFreeLagrangian}) and (\ref{EBbracketANDHam}), where $\mathbf{E}$ and $\mathbf{B}$ are the (non-canonical)  dynamic variables. Thus we can conclude that in the context of the non-local formulation exposed here, electric-magnetic duality is analogous to the phase space rotation symmetry of the SHO, with the conserved quantity being 
\begin{equation}\label{dualityCharge}
Q_\text{D} = \frac{1}{2}\iint d^3xd^3x'\omega_{ij}(\mathbf{x},\mathbf{x}')\left[E^i(x)E^j(x')+B^i(x)B^j(x')\right].
\end{equation}

Assuming now that the electric and magnetic fields are transverse, the vector potentials $\mathbf{A}(x)$ and $\mathbf{Z}(x)$ can be introduced such that $\mathbf{E} = - \nabla \times \mathbf{Z}$ and $\mathbf{B} = \nabla \times \mathbf{A}$. It is then easily proven that the above non-local quantity (\ref{dualityCharge}) becomes the local
\be
Q_\text{D} = \frac{1}{2}\int d^3x [\mathbf{Z}\cdot(\nabla\times\mathbf{Z}) + \mathbf{A}\cdot(\nabla\times\mathbf{A})] \ ,
\ee
equivalent to the conserved charge obtained by Calkin \cite{calkin1965invariance} and Deser-Teitelboim \cite{deser1976duality}. An extended discussion in the quantum theory is given in \cite{agullo2017electromagnetic, agullo2017essay, agullo2018electromagnetic, agullo2018symmetry}.

 We would like to remark that the conservation law $\frac{d}{dt} Q_\text{D} =0$ should be modified in the presence of charged matter, since  duality rotations are no longer symmetries of the theory.  Note~that this is somewhat similar to the chirality transformation of fermions \cite{qftbook}. Chirality rotations are symmetries for massless fermions, implying that $\partial_\mu j_5^\mu=0$, where $j^\mu_5 \equiv \bar\psi \gamma^\mu\gamma^5 \psi $, and the corresponding conservation of the chiral charge $Q_5\equiv \int d^3x j^0_5$. In presence of a mass term, $\frac{d}{dt} Q_\text{5} =0$ would also be modified~accordingly.

\section{The Non-Local Action with Matter\label{nonlocalLagrangianSection}}
The non-local action presented in the previous section can be straightforwardly generalized  to accommodate for the presence of matter. This is a important issue since the interaction of the electromagnetic field with matter has both fundamental and applied significance.  This new action functional $S_\text{NL}[\mathbf{E},\mathbf{B},\lambda] = \int d^4x\mathcal{L}_\text{NL}$, essentially based 
on Ref. \cite{jackiw1993constrained}, has the electric and magnetic fields as its dynamical fields as well as a Lagrange multiplier $\lambda$ that imposes Gauss' law (\ref{gauss}) as a constraint,
\begin{equation}\label{nonlocalLagrangian}
\mathcal{L}_{\text{NL}}=\int d^3 x' \ \omega_{ij}(\mathbf{x},\mathbf{x} ')[\partial_tE^i(x) + J^i(x)]B^j(x') - \frac{1}{2}(\mathbf{E}^2 + \mathbf{B}^2) - \lambda(\nabla \cdot \mathbf{E} - \rho).
\end{equation}
In the above expression $\omega_{ij}(\mathbf{x},\mathbf{x} ')$ is again given by (\ref{symplectic}). We note that a single Lagrange multiplier $\lambda$ is introduced here, instead of the two employed in Ref. \cite{jackiw1993constrained}. This Lagrangian provides all four of Maxwell's equations if there is electric charge conservation, i.e., $\dot{\rho} + \nabla\cdot\mathbf{J} = 0$, a prerequisite that is used in the standard formulation (\ref{standardLagrangian}) as well to preserve gauge invariance.  For instance, if the matter field is  given by a Dirac spinor $\psi$, with electric charge $q$ and mass $m$, we should replace $\rho = q\bar \psi \psi$ and $J^i = q\bar \psi \gamma^i \psi$ in (\ref{nonlocalLagrangian}). One can then complete the action by adding the standard local free action for the Dirac field such that the Lagrangian of the complete theory reads

\begin{equation}\label{nonlocalLagrangian2}
         \mathcal{L} = (i\bar \psi \gamma^\mu \partial_\mu \psi - m\bar \psi \psi ) + \int d^3 x' \ \omega_{ij}(\mathbf{x},\mathbf{x} ')[\partial_tE^i(x) + q\bar \psi \gamma^i \psi(x)]B^j(x') - \frac{1}{2}(\mathbf{E}^2 + \mathbf{B}^2) - \lambda(\nabla \cdot \mathbf{E} - q\bar \psi \psi).
\end{equation}

In addition to the constraint (\ref{gauss}) enforced by $\lambda$, the equations of motion for the action (\ref{nonlocalLagrangian}) are
\begin{equation}\label{deltaE}
E^i = \partial_i \lambda -\int d^3x' \omega_{ij}(\mathbf{x},\mathbf{x}') \partial_t B^j (x') 
\end{equation}
\begin{equation}\label{deltaB}
B^i =\int d^3x' \omega_{ij}(\mathbf{x},\mathbf{x}') [\partial_t E^j (x') + J^j(x')] \ , 
\end{equation}
which correspond to the Helmholtz decomposition of the electromagnetic field coupled to an external source. Gauss' law for the magnetic field is recovered by taking the divergence of (\ref{deltaB}), while the time-dependent Maxwell Equations (\ref{ampere}) and (\ref{Mh1}) are obtained by applying a curl on (\ref{deltaB}) and~(\ref{deltaE})~respectively.

The standard formalism in terms of the potentials can also be recovered solving the non-time evolving Equation (\ref{Mh2}). Applying the variable change $\mathbf{B}\rightarrow \mathbf{A}$ such that $\mathbf{B} = \nabla \times \mathbf{A}$ along with the relabelling $A^0 \equiv -\lambda$, it can be shown that (\ref{nonlocalLagrangian}) becomes
\begin{equation}\label{preliminaryL}
\mathcal{L} = \left(\partial_t\mathbf{E}+\mathbf{J}\right)\cdot\mathbf{A}_\text{T} - \frac{1}{2}\left[\mathbf{E}^2 + (\mathbf{\nabla}\times\mathbf{A}_\text{T})^2\right] +A^0\left(\mathbf{\nabla}\cdot\mathbf{E} - \rho\right) \ 
\ , \end{equation}
 which is of a similar form to (\ref{EandAT}). Hence, the introduction of the vector potential makes the non-local Lagrangian density become the standard first-order Lagrangian density  after removing the excess longitudinal component of $\mathbf{A}$. However, it is important to keep in mind  that (\ref{nonlocalLagrangian}) and (\ref{preliminaryL})  are not fully equivalent, as  the equation $\mathbf{\nabla}\cdot\mathbf{B}=0$ holds as a proper Euler-Lagrange equation  for (\ref{nonlocalLagrangian}), while 
it is assumed off-shell for (\ref{preliminaryL}).

Nevertheless, it is not difficult to see that (\ref{nonlocalLagrangian}) can be obtained by introducing the explicit expression of $\mathbf{A}_\text{T}$ into (\ref{preliminaryL})
\begin{equation}\label{coulombA}
A^i_\text{T}(x) = \int d^3 x' \omega_{ij}(\mathbf{x},\mathbf{x}')B^j(x')
\end{equation}
and assuming (\ref{Mh2}) holds. Therefore, even though the formalism in terms of (\ref{nonlocalLagrangian}) is not equivalent to the one of (\ref{standardLagrangian}) or (\ref{preliminaryL}), in some instances it will be useful to obtain results for the non-local viewpoint by simply substituting (\ref{coulombA}) wherever $\mathbf{A}$ appears in results derived from the local viewpoint, which is equivalent to imposing the Coulomb gauge, i.e., $\nabla \cdot \mathbf{A}=0$ or $\mathbf{A} = \mathbf{A}_\text{T}$. 
This property can be illustrated by considering the Lagrangian of the non-relativistic particle (\ref{standardParticleLagrangian}). Inserting (\ref{coulombA}) and relabelling $\lambda \equiv - A^0$, a new Lagrangian is obtained,
\be\label{classicalEMparticle}
L_\text{NL,p}[\mathbf{x}] = \frac{1}{2}m\dot{\mathbf{x}}^2 + e\int d^3 x' \omega_{ij}(\mathbf{x},\mathbf{x}')\ \dot{x}^i B^j(x')+ e\lambda(x).
\ee

Alternatively, (\ref{classicalEMparticle}) could have been obtained by applying the same procedure that was used to obtain (\ref{standardParticleLagrangian}) on (\ref{nonlocalLagrangian}).
While the Lagrangian $L_\text{NL,p}$ appears to be non-local with respect to the magnetic field, the equations of motion are expectedly the Lorentz force (\ref{lorentzForce}), which is local in both $\mathbf{E}$ and $\mathbf{B}$. This~is reassuring, as in a classical $\delta S = 0$ context no possibly non-local phenomenon is observed. 

\begin{figure}[htbp]
\begin{center}
\includegraphics[width=10cm]{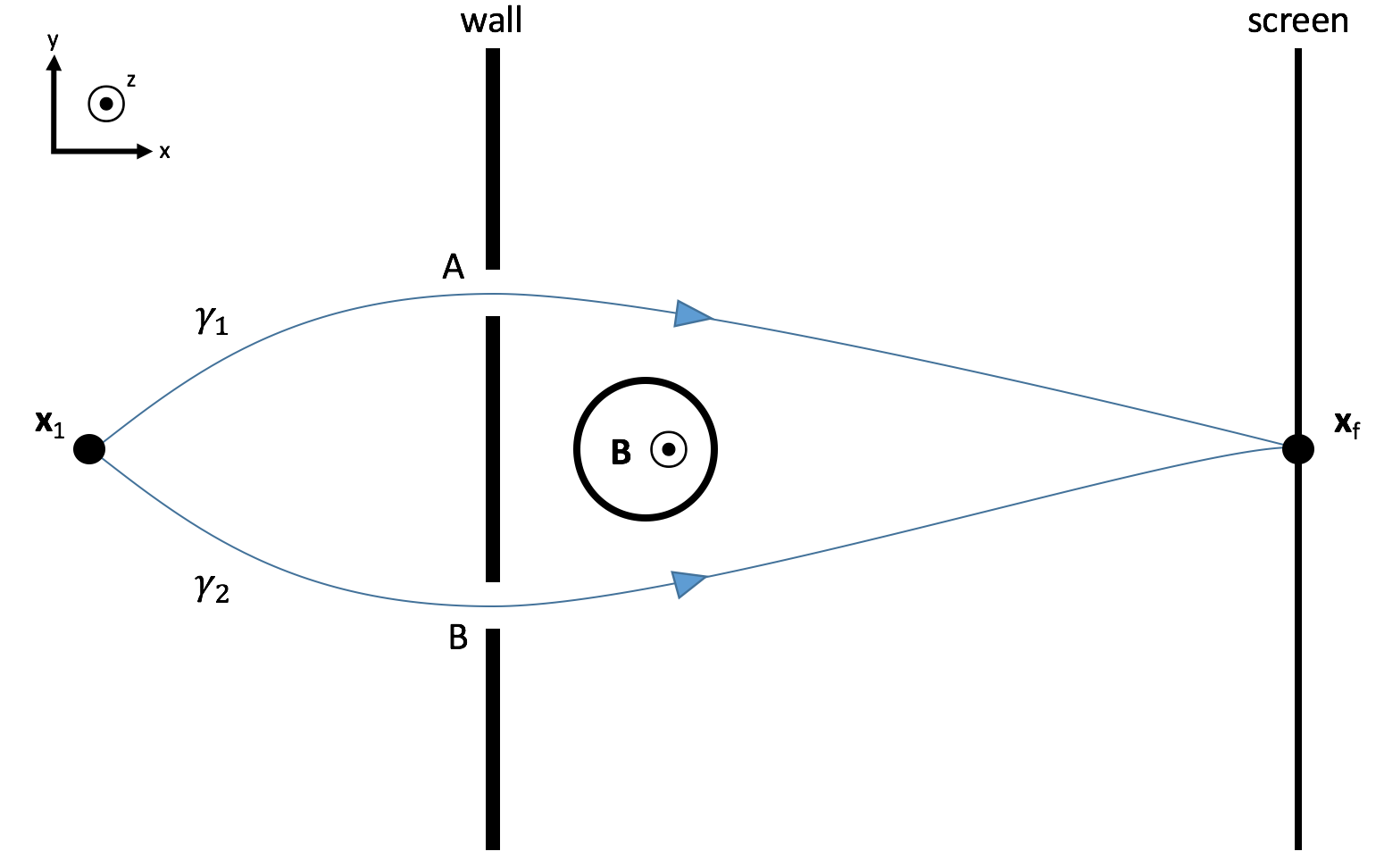}
\end{center}
\caption{\small{Experimental setup we will consider to analyse the magnetic AB effect. A source of electrons is located at the point $\mathbf{x}_1$, from which one is emitted at a time $t_1$. Between the source and a screen on the other side of the setup there is a wall, containing two slits A and B, and a long impenetrable cylinder behind it. Inside the cylinder, oriented parallel to the z-axis, there is a magnetic field $\mathbf{B}=\hat{\mathbf{z}}B_0$, while outside  $\mathbf{B} = 0$. The electrons can trace two types of $\textit{deterministic}$ paths to reach the point $\mathbf{x}_\text{f}$ on the screen at a time $t_\text{f}$, either above (e.g., $\gamma_1$) or below (e.g., $\gamma_2$) the cylinder.}}
\label{setupfigure}
\end{figure}

Things are not so simple however in a quantum context, a fact best depicted by considering the magnetic AB effect  with Feynman's path integral method. The details of the setup considered here to analyse the AB effect are described in Figure \ref{setupfigure}.
The action for this process is given by $S_\text{NL,p} = \int dt L_\text{NL,p}$ with $\lambda = 0$, and it can thus be proven that the propagator for the electrons getting from the source to the screen is
\be\label{nonlocalPaths}
K(\mathbf{x}_\text{f},t_\text{f};\mathbf{x}_\text{1},t_\text{1}) = \exp{\left[\frac{ie}{\hbar}\int_\text{above}d\text{s}^i\int d^3 x' \omega_{ij}(\mathbf{s},\mathbf{x}') B^j(x')\right]}\int_\text{above} \mathcal{D}[\mathbf{x}(t)]\exp\left[\frac{iS_0}{\hbar}\right]
\nonumber
\ee
\be
+ \exp{\left[\frac{ie}{\hbar}\int_\text{below}d\text{s}^i\int d^3 x' \omega_{ij}(\mathbf{s},\mathbf{x}') B^j(x')\right]}\int_\text{below} \mathcal{D}[\mathbf{x}(t)]\exp\left[\frac{iS_0}{\hbar}\right].
\ee

This result can be obtained using an analogous method to the one shown in Ref. \cite{sakurai2017modern}. The~term $S_0 = \int dt \frac{1}{2}m\dot{\mathbf{x}}^2$ is the free particle action while subscripts ``above'' and ``below'' in (\ref{nonlocalPaths}) are used to distinguish paths that curl above the cylinder from those that curl below. 
As it is known from the standard analysis of the AB effect, all paths curling above have a common phase, while those curling below have another, a property that appears explicitly in (\ref{nonlocalPaths}).
In contrast to the standard analysis however, these phases are explicitly non-local with respect to the physically relevant quantity,
the~magnetic field $\mathbf{B}$ inside the cylinder, instead of being local in the vector potential $\mathbf{A}$ outside.
Therefore, the non-locality suggested by the standard derivation of the magnetic AB effect appears naturally in the non-local prescription of electrodynamics described here. While the result (\ref{nonlocalPaths}) can be derived by simply applying the Coulomb gauge on (\ref{standardParticleLagrangian}) \cite{stewart2013role}, we stress how here it has really been proven from a more fundamental action (\ref{nonlocalLagrangian}), and not from an arbitrary choice of gauge.


The cylindrical symmetry of the setup ensures that an analytical value of the nonlocal interaction term, equivalent to the transverse component $\mathbf{A}_\text{T}$ of the vector potential (\ref{coulombA}), can be obtained,
\be\label{nlitAB}
\int d^3 x' \omega_{ij}(\mathbf{x},\mathbf{x}') B^j(x') = \left[\frac{\Phi_\mathbf{B}}{2\pi\rho}\hat{\pmb{\varphi}}\right]^i,
\ee 
where $\rho^2 = (x^1)^2 + (x^2)^2$ is the distance squared with respect to the center of the cylinder and $\hat{\pmb{\varphi}}$ is the unit vector associated with the azimuthal angle. This result can be derived by evaluating the volume integral directly as we have done for completeness in the Appendix, or treating $\mathbf{A}_\text{T}$ as a shorthand for the interaction term (left-hand-side (LHS) of (\ref{nlitAB})) and recycling the standard derivation ~\cite{sakurai2017modern}. The~relevant phase difference is thus the expected AB phase,
\be\label{phasedifference}
\Delta \varphi = \frac{e}{\hbar}\left[\int_\text{above}\mathbf{A}_\text{T}\cdot d\mathbf{s} - \int_\text{below}\mathbf{A}_\text{T}\cdot d\mathbf{s}\right] = \frac{e\Phi_\mathbf{B}}{\hbar}.
\ee
where $\Phi_\mathbf{B}$ is the magnetic flux through the cylinder.

\section{Conclusions}

Non-locality is a reasonably objectionable feature, but we feel the fomulation of electrodynamics treated here,  elaborating and improving on a proposal sketched in \cite{jackiw1993constrained},  will at least be useful to shed some light on the subtle topic of action functionals independent of potentials. We have argued how non-locality seems to be unescapable in an electromagnetic field-dependent formalism due to the non-trivial commutation relations $\{\mathbf{E},\mathbf{B}\}$. It is nonetheless important to keep in mind that the field-matter action (\ref{nonlocalLagrangian}) is not completely independent of potentials, as the Lagrange multiplier $\lambda$  in (\ref{nonlocalLagrangian}) is actually just a relabelled (Coulomb gauge) scalar potential.
However, it is consistent to assume $\lambda = 0$ in the context of electric-magnetic duality or the magnetic AB effect, meaning they can be studied without concern. 

On one hand, the former can be seen as a manifestation of the phase-space rotation symmetry of the SHO. It is worth recalling how this symmetry was derived with an action where all the Maxwell equations hold solely on-shell, in contrast with past derivations, which assume some of them off-shell. On the other hand, an arguably plausible interpretation for the AB effect was deduced. In a classical context, where $\delta S = 0$, the equations of motion (\ref{lorentzForce}) of (\ref{classicalEMparticle}) are local in both $\mathbf{E}$ and $\mathbf{B}$ despite the non-locality of the action. Therefore the correspondence principle holds, i.e., when $\hbar \rightarrow 0$ the interaction of the particle with the electromagnetic field is local. In a quantum context however trajectories with $\delta S \neq 0$ are not negligible, hence the non-locality of the action can materialize (\ref{nonlocalPaths}) with the AB effect. Through this scope, manifest non-locality is thus an exclusively quantum affair, and we believe this is also one of the lessons of this note.

 We would like to remark that we are not advocating to avoid the use of field potentials to analyze electrodynamics or its generalizations (nonabelian gauge theories). The purpose of this work is to point out that it could be useful to reanalyze electrodynamics from a nonlocal perspective (using only the electric and magnetic fields).  In so doing this we have filled a gap in the literature and  obtain, as a bonus,  new insights on two important topics in electrodynamics: i) the electromagnetic duality symmetry, and ii) the AB effect.  

After finishing this work we became aware of the work \cite{Majumdar-Ray}, concerning a formulation of electrodynamics without a gauge-fixing procedure. We think that there is a close connection with our work that could merit to be further explored.

\vspace{6pt}

\appendix 



\section{Interaction Term in the A.B. Effect\label{interactionTermDerivation}}

Preliminary considerations:
\begin{itemize}
\item The expression for the magnetic field is $\mathbf{B}(\mathbf{x}) = \hat{\mathbf{z}}\Theta(R^2-x^2-y^2)$, where $\Theta$ is the Heaviside step function and $\mathbf{x} = (x,y,z)$.
\item The volume region is a cylinder $C$ of radius $R$, with a length $L_1$ and $L_2$ over and under the xy plane respectively. Furthermore, it will be assumed that the cylinder is long i.e., $L^2_1,L^2_2 \gg R^2,x^2+y^2 $.
\end{itemize}

Due to its equivalence with the nonlocal interaction term (LHS of  (\ref{nlitAB})), we will use $\mathbf{A}_\text{T}$ as a shorthand to refer to it. It can thus be proven that
\begin{equation}
\mathbf{A}_\text{T}(\mathbf{x})
= \frac{B_0}{4\pi}\hat{\mathbf{z}}\times\int_C d^3x'\nabla\left(\frac{1}{|\mathbf{x} - \mathbf{x}'|}\right)
= \frac{B_0}{4\pi}\hat{\mathbf{z}}\times\int_{\partial C} d\mathbf{S}'\ \frac{1}{|\mathbf{x} - \mathbf{x}'|}.
\nonumber
\end{equation}
where a corollary of the Divergence theorem was used in the second equality.

The surface of the cylinder is composed by a circular wall and the two lids on either end. However,~since the lids have a normal vector $d\mathbf{S}'\propto \hat{\mathbf{z}}$ and $\hat{\mathbf{z}}\times\hat{\mathbf{z}}=0$, their contributions to the total integral are 0. Consequently, the only relevant contribution to the integral comes from the circular wall, with a normal vector $d\mathbf{S}' = \hat{\pmb{\rho}}'Rd\phi'dz'$ where $\hat{\pmb{\rho}}' = (\cos\phi',\sin\phi',0)$:

\begin{equation}
= \frac{B_0}{4\pi}\hat{\mathbf{z}}\times\int^{2\pi}_0 Rd\phi'\hat{\pmb{\rho}}'\int^{L_1}_{-L_2}dz'\ \left({z'}^2 + \alpha(\phi')\right)^{-1/2}
\end{equation}
\begin{equation}
= \frac{B_0}{4\pi}\hat{\mathbf{z}}\times\int^{2\pi}_0 Rd\phi'\hat{\pmb{\rho}}'
\left[\log\left(\sqrt{\alpha(\phi')+L^2_1} + L_1\right)+\log\left(\sqrt{\alpha(\phi')+L^2_2} + L_2\right) - \log\left(\alpha(\phi')\right)\right]
\nonumber
\end{equation} \\
where $\alpha(\phi') = (x-R\cos\phi')^2 + (y-R\sin\phi')^2$ was introduced for brevity.
However,~expressions~of the form $\log\left(\sqrt{\alpha(\phi')+L^2} + L\right)$ can be disregarded by taking into account the first preliminary consideration, 
\begin{equation}
\int^{2\pi}_0d\phi'\hat{\pmb{\rho}}'\log\left(\sqrt{\alpha(\phi')+L^2} + L\right) \approx \log(2L)\int^{2\pi}_0d\phi'\hat{\pmb{\rho}}'= 0.
\nonumber
\end{equation}

\begin{figure}[htbp]
\includegraphics[width=15 cm]{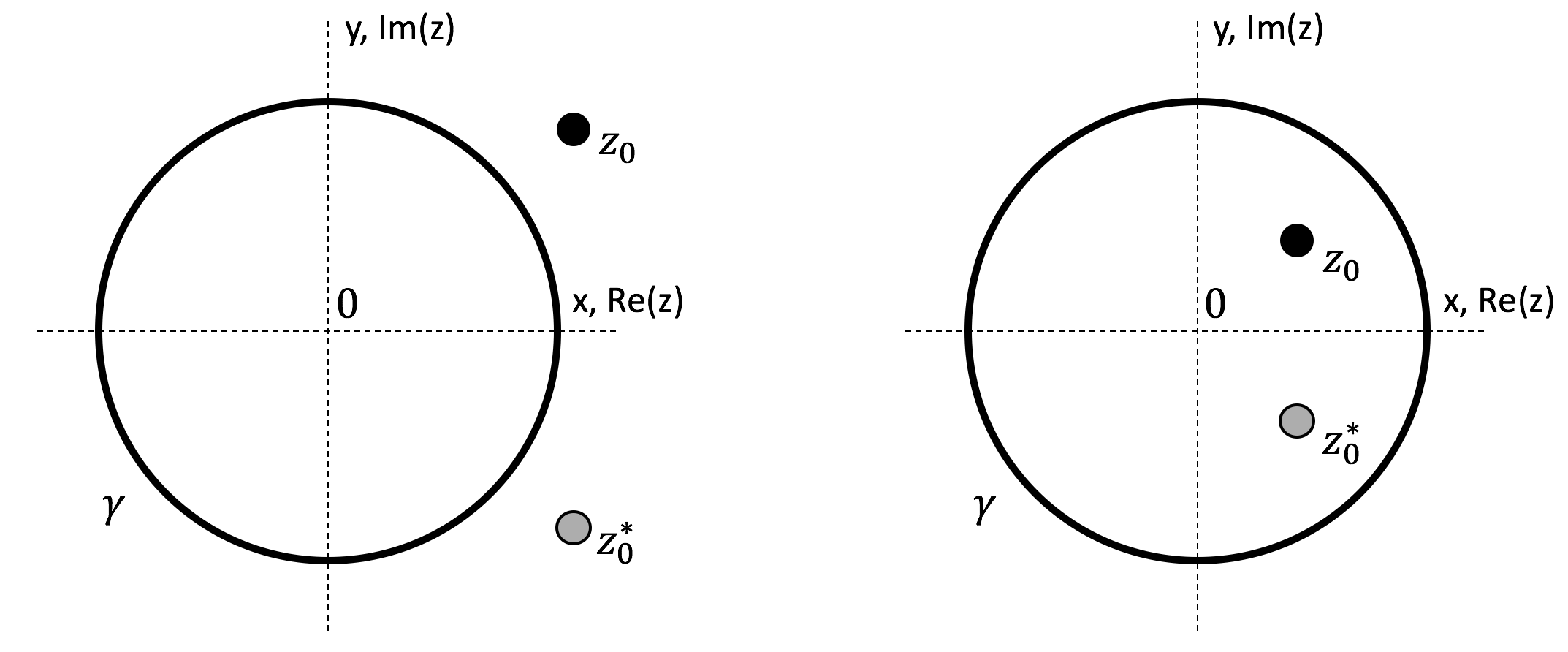}
\centering
\caption{Complex plane representation of $z_0 = x + iy$ outside and inside the cylinder.}\label{cplane} 
\end{figure}

Therefore the expression for $\mathbf{A}_\text{T}$ is now a one-dimensional integral
\be\label{A1D}
\mathbf{A}_\text{T}(\mathbf{x}) = -\frac{B_0R}{4\pi}\int^{2\pi}_0 d\phi'\hat{\pmb{\phi}}'\log\left[(x-R\cos\phi')^2 + (y-R\sin\phi')^2\right].
\ee
where $\hat{\mathbf{z}}\times\hat{\pmb{\rho}}'=\hat{\pmb{\phi}}'$, with $\hat{\pmb{\phi}}'=(-\sin\phi',\cos\phi',0)$. Equation (\ref{A1D}) can be reinterpreted as a complex integral, $z_\mathbf{A} = A_\text{T}^x +i A_\text{T}^y$, over a circle of radius $R$ on the complex plane
\be
z_\mathbf{A}(x,y) =-\frac{B_0}{4\pi}\oint_\gamma dz \log|z-z_0|^2
\ee
where $z_0 = x+iy$ and $\gamma(s) = Re^{is}$. Ignoring for now the multiplicative constant $-B_0/4\pi$, the integral can be split into two, 
\be\label{complex}
\oint_\gamma dz \log|z-z_0|^2=\oint_\gamma dz \log(z-z_0)+\oint_\gamma dz \log(z^*-z^*_0)
\ee
\be\label{complex2}
=\oint_\gamma dz \log(z-z_0)+R^2\oint_\gamma dz \frac{\log(z-z^*_0)}{z^2}
\ee
where the latter equality is due to the easily proven general property for circular contour integrals,
$\oint_\gamma dz \ f(z^*) = R^2\oint_\gamma dz\  f(z)/z^2$. The value of (\ref{complex2}) will depend on whether $z_0$ is inside or outside the disk delimited by $\gamma$ on the complex plane $\mathbb{C}$ (see Figure \ref{cplane}). In terms of the original problem, this~means that the expressions for $\mathbf{A}_\text{T}$ inside and outside the solenoid will be different. In the latter case, $\rho^2 \equiv x^2+y^2 > R^2$, meaning that 
\be
z_\mathbf{A}(x,y) = \frac{B_0\pi R^2}{2\pi} \frac{-y+ix}{x^2+y^2}.
\ee
and
\be
\mathbf{A}_\text{T}(\mathbf{x}) = \frac{B_0\pi R^2}{2\pi(x^2+y^2)}(-y,x,0) = \frac{B_0\pi R^2}{2\pi\rho}\hat{\pmb{\phi}} = \frac{\Phi_\mathbf{B}}{2\pi\rho}\hat{\pmb{\phi}}
\ee
where $\Phi_\mathbf{B} = B_0\pi R^2$ and $\hat{\pmb{\phi}}=(-y/\rho,x/\rho,0)$. This is the expected result outside the cylinder. On the other hand, inside $\rho^2 \equiv x^2+y^2 < R^2$. Evaluating (\ref{complex2}) in this situation gives
\be
z_\mathbf{A} = \frac{B_0}{2}(-y+ix)
\ee
so that
\be
\mathbf{A}_\text{T}(\mathbf{x}) = \frac{B_0}{2}(-y,x,0) = \frac{B_0\rho}{2}\hat{\pmb{\phi}}.
\ee

This is the expected result for the transverse component of the vector potential in a finite volume under a constant magnetic field (in this case, a cylinder with $\mathbf{B}=B_0\hat{\mathbf{z}}$), where $\mathbf{A}_\text{T}(\mathbf{x}) = -\frac{1}{2} \mathbf{x}\times \mathbf{B}$.

\acknowledgments{ J. N.-S. thanks  I. Agullo and A. del Rio  for useful~discussions.This research was funded with  Grants.  No.\  FIS2017-84440-C2-1-P, No. \ FIS2017-91161-EXP, No. \  SEV-2014-0398, and No. \ SEJI/2017/042 (Generalitat Valenciana). 
}

\end{document}